\documentstyle[12pt]{article}

\begin{document}
\setcounter{page}{1}
\renewcommand{\thefootnote}{\fnsymbol{footnote}}
\pagestyle{plain} \vspace{1cm}
\begin{center}
\Large{\bf Noncommutative Inspired Reissner-Nordstr\"{o}m Black
Holes in Large Extra Dimensions}
\\
\small \vspace{1cm} {\bf Kourosh Nozari$^{\rm
a,b,}$\footnote{knozari@umz.ac.ir}}\quad\quad and \quad\quad {\bf S.
Hamid Mehdipour$^{\rm c,}$\footnote{
mehdipour@iau-lahijan.ac.ir}}  \\
\vspace{0.5cm} {\it $^{a}$Department of Physics, Faculty of Basic
Sciences,
University of Mazandaran,\\
P. O. Box 47416-1467, Babolsar, IRAN}\\
{\it $^{b}$Research Institute for Astronomy and Astrophysics of
Maragha, \\P. O. Box 55134-441, Maragha, IRAN }\\
{\it $^{c}$ Islamic Azad University, Lahijan Branch,\\
P. O. Box 1616, Lahijan, IRAN} \\
\end{center}
\vspace{1.5cm}
\begin{abstract}
Recently, a new noncommutative geometry inspired solution of the
coupled Einstein-Maxwell field equations including black holes in
4-dimension is found. In this paper, we generalize some aspects of
this model to the Reissner-Nordstr\"{o}m (RN) like geometries with
large extra dimensions. We discuss Hawking radiation process based
on noncommutative inspired solutions. In this framework, existence
of black hole remnant and
possibility of its detection in LHC are investigated.
\end{abstract}
\newpage

\section{ Introduction}
The underlying physics of black holes have been the target of many
investigations. One of the important characteristic of a black hole
is its thermodynamical properties: a black hole has Hawking
temperature [1] which is proportional to its surface gravity on the
horizon, and entropy of which is proportional to its horizon area
[2]. These two quantities satisfy the first law of black hole
thermodynamics. In this regard, studying thermal properties of
various black holes is one of the significant subjects of black hole
physics. Hawking has interpreted the quantum effect of black hole
emission as a thermal radiant spectrum from event horizon, which
sets a significant event in black hole physics. The discovery of
this effect solved and revealed both the problem in black hole
thermodynamics and the relation between quantum gravity and
thermodynamics. Hawking has pointed out that when the virtual
particles near the surface of the black hole with negative energy
come into black hole via tunnel effect, the energy of the black hole
will decrease and the radius of the black hole event horizon will
decrease also. This process is equivalent to the emission of a
particle from the black hole (black hole evaporation). But, how is
the final stage of black hole evaporation? The final stage of the
black hole evaporation is a matter of debates in the existing
literature [3]. The generalized uncertainty principle (GUP),
motivated by string theory and noncommutative quantum mechanics,
suggests significant modifications to the Hawking temperature and
evaporation process of black holes. Adler {\it et al} [4] have
argued that contrary to standard view point, GUP may prevent small
black holes from total evaporation in exactly the same manner that
the usual uncertainty principle prevents the Hydrogen atom
from total collapse.\\
Nicolini, Smailagic and Spallucci (NSS) [5] have found a
noncommutative geometry inspired solution of the Einstein equation
smoothly interpolating between a de Sitter core around the origin
and an ordinary Schwarzschild spacetime at large distances. Many
studies have been performed in these directions where spacetime is
commutative. Noncommutative spacetime view point [6], gets special
appeal due to telling beforehand of string theory, leads to the fact
that spacetime points might be noncommutative. Undoubtedly,
spacetime noncommutativity can cure some kind of divergences, which
appear in General Relativity. The inclusion of noncommutativity in
black hole metric has been studied in [7,8]. It has been shown that
the modified metric due to noncommutativity of spacetime does not
allow the black hole to decay beyond a {\it minimal mass} $M_0$.
Then, the evaporation process terminates when black hole reaches a
Planck size remnant with zero temperature, which does not diverge at
all, rather it reaches a maximum value before cooling
down to absolute zero.\\
The authors in Ref.~[9] have generalized the NSS model to the case
where flat, toroidally compactified extra dimensions are accessible
at the $TeV$ energy scale. During the last decade, several models
using compactified large extra dimensions (LEDs) scenarios [10,11]
have been proposed, which have significant implications for
processes involving strong gravitational fields, such as the decay
of black holes. In models with extra spatial dimensions the four
dimensional spacetime is viewed as a $D_3$-brane embedded in a bulk
spacetime of dimension $d$, (where $d\geq4$). Embedding a black hole
in a spacetime with additional dimensions would seem, from the
string theory point of view, to be a natural thing to do. For
extra-dimensional gravity with $TeV$ energy scales, Hawking
temperature and evaporation process of black holes lead to important
changes in the formation and detection of black holes at the Large
Hadronic Collider (LHC) [13]. Since a black hole can evaporate into
all existing particles whose masses are lower than its temperature,
thus these fascinating processes could be tested at the LHC, and
providing a way of testing the existence of extra dimensions.\\
Recently, Ansoldi, Nicolini, Smailagic and Spallucci (ANSS) [14]
along their previous studies, have found a new, noncommutative
inspired solution of the coupled Einstein-Maxwell field equations
including black holes in 4-dimensional brane universe. In this paper
we are going to generalize their model to large extra dimensions
scenario. So, the main purpose of this paper is to consider the
effect of space noncommutativity on the short distance
thermodynamics of an evaporating RN black hole in $d$-dimensional
spacetime. We investigate the possibility of formation of black
holes remnants and we discuss the energy scales for detection of
these remnants at LHC. We also discuss the evidences for
non-extensive thermodynamics of such a short distance systems.

The layout of the paper is as follows: we begin in Section $2$ by
outlining the RN black holes in spacetime dimensions higher than
four and their generalizations to a regular de Sitter vacuum
accounting for the effect of noncommutative coordinate fluctuations
at short distances (noncommutative inspired RN-dS solutions) in
$d$-dimensional bulk spacetime. In Section $3$ we pay special
attention to the thermodynamic behavior of RN-dS black holes by
study of Hawking temperature, entropy, specific heat and free energy
in various dimensions. The paper follows by summary and discussion
in Section $4$.

\section{ Noncommutative Inspired Charged Black Holes in Large Extra Dimensions}
The RN black hole is a solution of the Einstein equation coupled to
the Maxwell field. The classical RN metric is
\begin{equation}
ds^2=\frac{\Delta}{r^2}dt^2-\frac{r^2}{\Delta}dr^2-r^2d\Omega_2^2,
\end{equation}
where $d\Omega_2^2$ is the metric on the unit $S^2$ and
\begin{equation}
\Delta\equiv r^2-2Mr+Q^2\equiv(r-r_+)(r-r_-),
\end{equation}
with
\begin{equation}
r_{\pm}=M\pm\sqrt{M^2-Q^2}.
\end{equation}

Let us now consider the charged black hole thermodynamics in model
universes with large extra dimensions. There are two main scenarios
of large extra dimensions (LEDs)\footnote{The model proposed by
Dvali, Gabadadze and Porrati (DGP) [12] is essentially different
with above mentioned scenarios since it predicts deviations from the
standard 4-dimensional gravity even over large distances. However,
in this paper we restrict our study to the ADD model.}
\begin{itemize}
\item
the Arkani-Hamed--Dimopoulos--Dvali (ADD) model [10], where the
extra dimensions are compactified toroidally and all of radius $R$.
This model was motivated by the desire to provide a solution to the
so-called hierarchy problem, that is, the sixteen orders of
magnitude difference between the electroweak energy scale and the
Planck scale;\\ and
\item
the Randall--Sundrum (RS) model [11], where the extra dimensions
have an infinite extension but are warped by a non-vanishing
cosmological constant. This model also solve the hierarchy problem
despite a different approach to the ADD model.
\end{itemize}
In LEDs scenario, RN metric can be written as follows

\begin{equation} ds^2 = \bigg(1 - \frac{2m}{r^{d-3}} +
\frac{q^2}{r^{2(d-3)}}\bigg) dt^2 - \bigg(1 - \frac{2m}{r^{d-3}} +
\frac{q^2}{r^{2(d-3)}}\bigg)^{-1} dr^2 - r^2 d\Omega^2_{(d-2)},
\end{equation}
where $d\Omega^2_{(d-2)}$ is the line element on the
$(d-2)$-dimensional unit sphere and $d$ is spacetime dimensionality.
The volume of the $(d-2)$-dimensional unit sphere is given by
\begin{equation} \Omega_{(d-2)} =
\frac{2\pi^{\frac{d-1}{2}}}{\Gamma(\frac{d-1}{2})}.
\end{equation}
$g_{00}$ is a function of mass and charge given in terms of
parameters $m$ and $q$ as follows
\begin{equation} m = \frac{8\pi G_d
}{(d-2) \Omega_{(d-2)}}M,
\end{equation}
and
\begin{equation}
q = \sqrt{\frac{8\pi G_d}{(d-2)(d-3)}} \:Q.
\end{equation}
$G_d$ is gravitational constant in $d$-dimensional spacetime which
in ADD model is given by
\begin{equation}
G_{d} = \frac{(2\pi)^{d-4}}{\Omega_{d-2}}M_{Pl}^{2-d},
\end{equation}
where $M_{Pl}$ is the $d$-dimensional Planck mass and there is an
effective 4-dimensional Newton constant related to $M_{Pl}$ by
\begin{equation}
M_{Pl}^{2-d}=4\pi G_{4}R^{d-4},
\end{equation}
where $R$ is the size of extra dimensions. It is necessary to note
that in this work, the conventions for definition of the fundamental
Planck scale $M_{Pl}$ are the same as which have been used by ADD
and also GT [15]. ( Hereafter we set the fundamental constants equal
to unity; $\hbar = c = k_B = 4\pi\epsilon_0 = 1$ ). In this section,
we will obtain and investigate the noncommutative inspired RN
solution for a black hole in large extra dimensions, where
noncommutativity can be taken as the correction to the RN black hole
metric and goes to zero when the strength of noncommutativity goes
to zero. The simplest noncommutativity that one can postulate is the
commutation relation $ [\, \mathbf{x}^i\ , \mathbf{x}^j\,]= i \,
\theta^{ij} $,\,  with a parameter $\theta$ which measures the
amount of coordinate noncommutativity in the coordinate coherent
states approach [16] and \, $\theta^{ij}$ \, is an antisymmetric
(constant) tensor of
dimension $(length)^2$.\\
The approach we adopt here is to look for a static, asymptotically
flat, spherically symmetric, minimal width, Gaussian distribution of
mass and charge whose noncommutative size is determined by the
parameter $\sqrt{\theta}$. To do this end, we shall model the mass
and charge distributions by a smeared delta function $\rho$
([5,9,14])

\begin{displaymath}
\left\{ \begin{array}{ll}
 \rho_{matt}(r)={M\over {(4\pi \theta)^{\frac{d-1}{2}}}}
e^{-\frac{r^2}{4\theta}}\\
\\
 \rho_{el}(r)={Q\over {(4\pi \theta)^{\frac{d-1}{2}}}}
e^{-\frac{r^2}{4\theta}}.\\
\end{array} \right.
\end{displaymath}
The assumption of spherical symmetry means that the line element
reduces to the canonical form, namely,
\begin{equation}
ds^2=e^\nu dx_0^2-e^\mu dr^2-r^2 d\Omega_{d-2}^2\,,
\end{equation}
and
$$
d\Omega_{d-2}^2=d\vartheta^2_{d-3} + \sin^2\vartheta_{d-3}
\,\biggl(d\vartheta_{d-4}^2 + \sin^2\vartheta_{d-4}\,\Bigl(\,... +
\sin^2\vartheta_2\,(d\vartheta_1^2 + \sin^2 \vartheta_1
\,d\varphi^2)\,...\,\Bigr)\biggr),
$$
where $0 <\varphi < 2 \pi$ and $0< \vartheta_i < \pi$, for $i=1,
..., d-3$. In the above formulae, $\nu$ and $\mu$ are functions of
$r$ only, because we impose the condition that the solution is
static and our assumption that the solution is asymptotically flat
requires: $\nu,\mu\to 0$ as $r \to \infty$; this will require that
$\nu=-\mu$ in the solutions of Einstein-Maxwell field equations.\\
The system of Einstein-Maxwell field equations is as follows
\begin{displaymath}
\left\{ \begin{array}{ll} R^B{}_A-\frac{1}{2}\, \delta^B{}_A\, R =
8\pi G_d\,\left(\, T^B{}_A\vert_{matt} +
T^B{}_A\vert_{el}  \,\right)\\
\\
\frac{1}{\sqrt{-g}}\, \partial_B\,\left(\, \sqrt{-g}\,
F^{BA}\, \right)= J^A,\\
\end{array} \right.
\end{displaymath}
where $T^B{}_A\vert_{matt}=diagonal\,
(-\rho_{matt}(r),\,\,p_r,\,\,p_{\vartheta_1},...,\,\,p_{\vartheta_{d-3}},\,\,p_\phi)$,
are comprised of a radial pressure $p_r=-\rho_{matt}(r)$ and
tangential pressure of a self-gravitating anisotropic fluid
$p_{\vartheta_i}=p_\phi=-\rho_{matt}(r)-\frac{r}{(d-2)}\partial_r\rho_{matt}(r)$,
while the electromagnetic energy-momentum tensor must take on the
form
\begin{displaymath}
F^{BA}=\delta^{0[\, B\,\vert}  \delta^{r\,\vert\, A \,]}\,
E_d\left(\, r\,;\theta\right)=\,E_d\left(\, r\,;\theta\right) \left(
\begin{array}{ccccc}
0 & -1 & 0 & \ldots & 0\\
1 & 0 & 0 & \ldots & 0 \\
0 & 0 & 0 & \ldots & 0 \\
\vdots & \vdots & \vdots & \ddots & \vdots \\
0 & 0 & 0 & \ldots & 0 \\
\end{array} \right),
\end{displaymath}
where smearing of the electric field reads
\begin{equation}
E_d\left(\
r\,;\theta\right)=\frac{Q}{r^{2(d-3)}}\,\frac{\gamma\left(\frac{d-1}{2},\frac{r^2}{4\theta}\right)}{\Gamma(\frac{d-1}{2})}.
\end{equation}
Then the Einstein field equations $G_{BA} = 8\pi G_d T_{BA}$ lead to
the following solution
\begin{equation}
ds^2\equiv g_{BA}dx^B\,dx^A= g_{00}\,dt^2 - g_{00}^{-1}\,dr^2 - r^2
d\Omega^2_{(d-2)},
\end{equation}
with
\begin{displaymath}
\left\{ \begin{array}{ll} g_{00}=1 -
\frac{2m}{r^{d-3}}\,\frac{1}{\Gamma(\frac{d-1}{2})}\gamma\left(\frac{d-1}{2},\frac{r^2}{4\theta}\right)
+\frac{(d-3)^2(d-2)}{2\pi^{d-3}}\frac{q^2}{r^{2(d-3)}}F(r)\\
F(r)=\gamma^2\left(\frac{d-3}{2},\frac{r^2}{4\theta}\right)-\frac{2^{\frac{11-3d}{2}}}{(d-3)\theta^{\frac{d-3}{2}}}
\gamma\left(\frac{d-3}{2},\frac{r^2}{4\theta}\right)r^{d-3}\\
\gamma\left(\frac{a}{b},u\right)=\int_0^u\frac{dt}{t}t^{\frac{a}{b}}e^{-t}.\\
\end{array} \right.
\end{displaymath}
In fact, by plugging the above metric into the Einstein-Maxwell
system, the $g_{00}$ can be determined, although it is done slightly
simpler, for the larger values of $d$ with a good approximation, by
plugging the explicit form for the smeared mass and charge
distributions into the metric as follows
\begin{equation}
g_{00}=1 - \frac{2m_\theta}{r^{d-3}}+ \frac{q_\theta^2}{r^{2(d-3)}},
\end{equation}
with
\begin{displaymath}
\left\{ \begin{array}{ll}
 m_\theta = \frac{8\pi G_d }{(d-2)
\Omega_{(d-2)}}M_\theta\\
\\
q_\theta = \sqrt{\frac{8\pi G_d}{(d-2)(d-3)}} \:Q_\theta,\\
\end{array} \right.
\end{displaymath}
where $M_\theta$ and $Q_\theta$ are the smeared mass and charge
distributions respectively and are decided by
\begin{displaymath}
\left\{ \begin{array}{ll}
M_\theta=\int_0^r\rho_{matt}(r)\Omega_{(d-2)}
r^2dr=\frac{\gamma\left(\frac{d-1}{2},\frac{r^2}{4\theta}\right)}{\Gamma(\frac{d-1}{2})}\,M\\
\\
Q_\theta=\int_0^r\rho_{el}(r)\Omega_{(d-2)}
r^2dr=\frac{\gamma\left(\frac{d-1}{2},\frac{r^2}{4\theta}\right)}{\Gamma(\frac{d-1}{2})}\,Q.\\
\end{array} \right.
\end{displaymath}
The above metric smoothly interpolates between de Sitter core around
the origin and an ordinary Reissner-Nordstr\"{o}m geometry far away
from the origin (RN-dS Black Hole in large extra dimensions). On the
other hand, the curvature singularity at $r=0$ is eliminated by
noncommutativity as an intrinsic property of the manifold. In this
situation, a regular de Sitter vacuum state will be formed
accounting for the effect of noncommutative coordinate fluctuations
at short distances and also a usual Reissner-Nordstr\"{o}m spacetime
at large distances is being taken into account in higher than
4-dimension. Classical RN metric for large distances in 4-dimension
is also obtained from (13) in the limit of \,$\theta\rightarrow0$\,
or \,
$r\gg\theta$\, and $d=4$. \\
The event horizon radius,\, $r_H$,\, can be obtained from the
equation $g_{00}\left(\, r_H\,\right)=0$ which gives
\begin{equation}
1 - \frac{2m_\theta}{r_H^{d-3}}+ \frac{q_\theta^2}{r_H^{2(d-3)}}=0.
\end{equation}
Depending on the different values of $Q$, $M$ and $M_{Pl}$, the
metric displays three possible causal structure [5,9,14]:
\textbf{1}- It is possible to have two distinct horizons
(Non-extremal Black Hole), \textbf{2}- It is possible to have one
degenerate horizon (Extremal Black Hole), and finally \textbf{3}- It
is impossible to have horizon at all (Massive Charged Droplet).

It is important to note that, the $d$-dimensional Planck mass
$M_{Pl}$ in LEDs models might be as low as a $TeV$-scale, because it
is found that, this $TeV$-scale is very directly constrained by
experimental bounds and it is also required to be $\sim 1 \,TeV$ in
order to solve the hierarchy problem, which is relevant for black
hole production at near-future experiments (LHC and also in ultra
high energy cosmic ray showers [17]). Consequently, the minimum
energy for possible formation and detection of black holes at LHC is
decreased, if $M_{Pl} \sim 1\, TeV$. Indeed, the minimal mass of
black hole depends sensitively on the fundamental Planck scale,
$M_{Pl}$, and on the spacetime dimension, $d$. Based on this
feature, in the following figures, $1$ and $2$, the fundamental
Planck mass has been set equal to $M_{Pl}=0.5\, TeV$, while in
figure $3$ we have set $M_{Pl}=1.5 \,TeV$. In all of these figures,
the initial mass of black hole has been chosen to be $M=5 \,TeV$.
These figures show that, if the initial mass of black hole as energy
scales accessible at the LHC is not large enough, then LHC will not
see any black hole in this regime. Also, figure $2$ shows that
possibility of black hole formation is reduced by increasing the
charge of black hole particularly for $4$-dimensional black hole on
the brane.

\begin{figure}[htp]
\begin{center}
\includegraphics{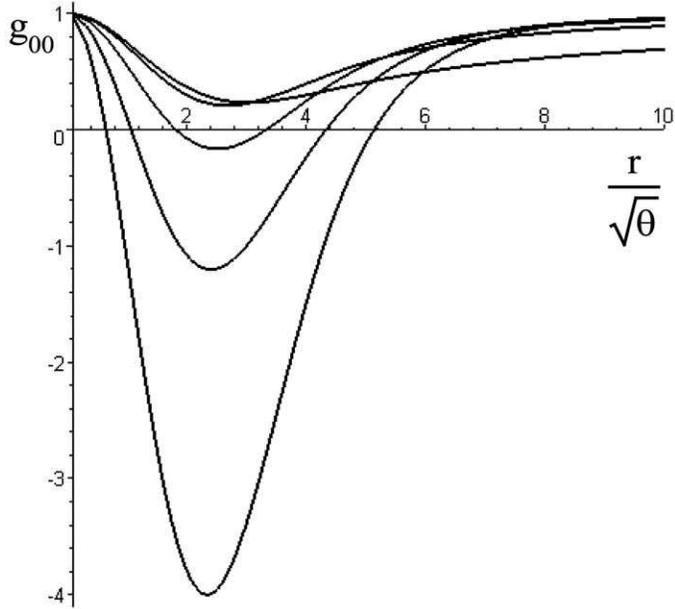}
\end{center}
\vspace{6.7 cm} \caption{\scriptsize {$g_{00}$ versus the radius \,
$r$ \, in $\sqrt{\theta}$ units for different dimensions. Black hole
charge, mass and $d$-dimensional Planck mass are set equal to
$Q=0.5$, $M=5$ and $M_{Pl}=0.5$ respectively. On the left-hand side
of the figure, curves are marked from top to bottom by $d = 4$ to $d
= 8$. This figure shows the possibility of having extremal
configuration by decreasing the number of spacetime dimensions. }}
\end{figure}

\begin{figure}[htp]
\begin{center}
\includegraphics{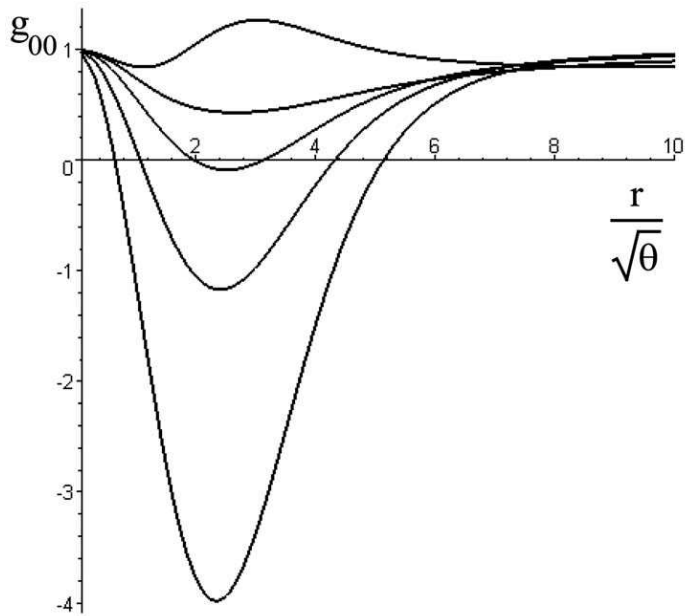}
\end{center}
\vspace{7 cm} \caption{\scriptsize {$g_{00}$ versus the radius\, $r$
\,in $\sqrt{\theta}$ units for different number of dimensions. Black
hole charge and mass and $d$-dimensional Planck mass are set equal
to $Q=2$, $M=5$ and $M_{Pl}=0.5$ respectively. On the left-hand side
of the figure, curves are marked from top to bottom by $d = 4$ to $d
= 8$. This figure is the same as previous one: possibility of
extremal configuration by decreasing the number of spacetime
dimensions. However, in comparison with previous figure, we see a
significant difference for black hole on the 3-brane when the charge
varies. This may be a reflection of the fact that black hole lives
on the brane and radiates mainly on the brane [18].}}
\end{figure}

\begin{figure}[htp]
\begin{center}
\includegraphics{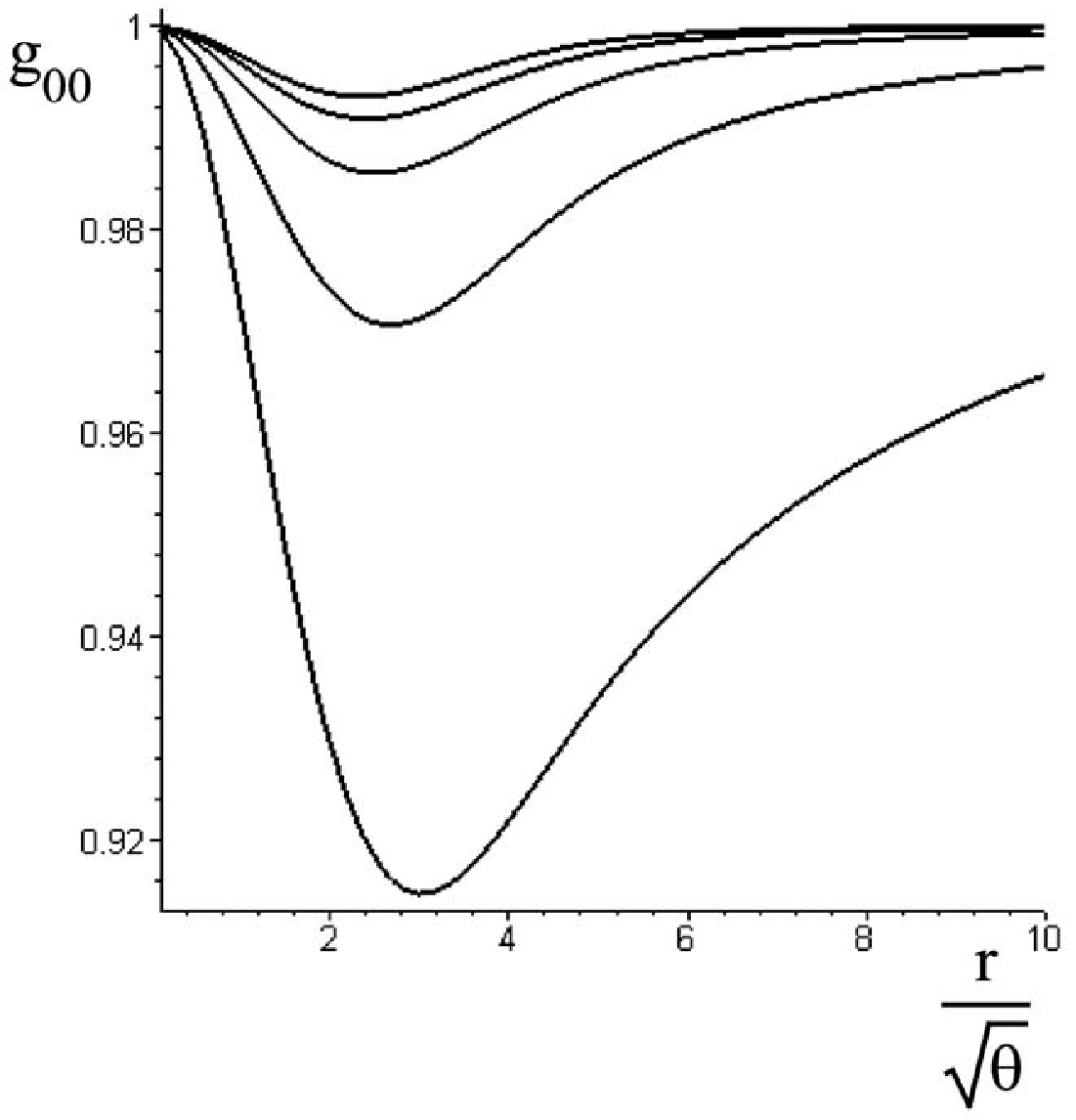}
\end{center}
\vspace{6.6 cm} \caption{\scriptsize {$g_{00}$ versus the radius,
$r$\, in $\sqrt{\theta}$ units for different number of spacetime
dimensions. Black hole charge and mass and the $d$-dimensional
Planck mass are set equal to $Q=0.5$, $M=5$ and $M_{Pl}=1.5$
respectively. On the left-hand side of the figure, curves are marked
from bottom to top by $d = 4$ to $d = 8$. The figure shows that in
this case there is no horizon and then no black hole is formed.}}
\end{figure}

Analytical solution of equation (14) for $r_{H}$ in a closed form is
impossible, so we solve it numerically to find this quantity.
However, it is possible to solve (14) to find $M$, which provides
the mass as a function of the horizon radius $r_H$ and charge $Q$ in
an arbitrary dimension. If we have chosen a finite dimension (for
example $d=4$, $d=5$ and so on), then the mass of RN-dS black hole
as a function of the horizon radius and charge can be obtained by
solving equation (14). This leads us to
\begin{equation}
d=4\Longrightarrow M=\frac{\sqrt{\pi}
r^2_H\theta+4G_4Q^2\bigg(\pi^{\frac{3}{2}}\theta\,
{\cal{E}}\Big(\frac{r_H}{\sqrt{2\theta}}\Big)^2
e^{\frac{r_H^2}{4\theta}}+\sqrt{\pi}r_H^2e^{-\frac{r_H^2}{4\theta}}-\frac{2\pi
r_H}{\sqrt{\theta}}\,{\cal{E}}\Big(\frac{r_H}{\sqrt{2\theta}}\Big)\bigg)}{-2G_4r_H^2\theta^{\frac{1}{2}}+2G_4r_H
\theta\sqrt{\pi}\,
\,{\cal{E}}\Big(\frac{r_H}{\sqrt{2\theta}}\Big)e^{\frac{r_H^2}{4\theta}}},
\end{equation}

\begin{equation}
d=5\Longrightarrow M=\frac{-\frac{3}{2}\pi
r_H^4\theta^2-\pi^2G_5e^{-\frac{r_H^2}{2\theta}}Q^2\bigg(\frac{r_H^4}{8}+r_H^2\theta-r_H^2\theta
e^{\frac{r_H^2}{4\theta}}+2\theta^2-4\theta^2e^{\frac{r_H^2}{4\theta}}+2\theta^2e^{\frac{r_H^2}{2\theta}}\bigg)}
{G_5r_H^4\theta
e^{-\frac{r_H^2}{4\theta}}+4G_5r_H^2\theta^2e^{-\frac{r_H^2}{4\theta}}-4G_5r_H^2\theta^2},
\end{equation}
and so on. ${\cal{E}}(x)$ shows the {\it Gauss Error Function}
defined as follows $$ {\cal{E}}(x)\equiv
\frac{2}{\sqrt{\pi}}\int_{0}^{x}e^{-t^2}dt.$$ When $d$ is even, we
see that these equations can be expressed in terms of combinations
of error functions. When $d$ is odd, it is possible to solve these
equations analytically. The results of numerical solution of the
mass of RN-dS black hole as a function of the horizon radius are
presented in figures $4$ and $5$.
\begin{figure}[htp]
\begin{center}
\includegraphics{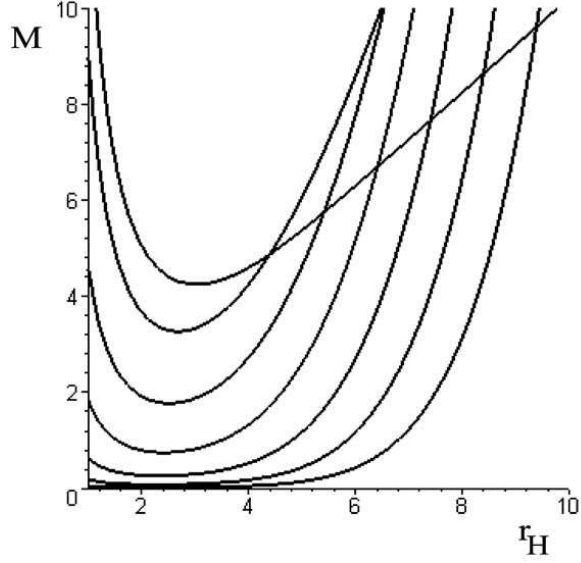}
\end{center}
\vspace{6  cm} \caption{\scriptsize {Black hole mass $M$ versus the
radius of event horizon, $r_H$, for different number of spacetime
dimensions. Black hole charge and the $d$-dimensional Planck mass
are set equal to $Q=0.5$ and $M_{Pl}=0.4$ respectively. On the
left-hand side of the figure, curves are marked from top to bottom
by $d = 4$ to $d = 10$. Since the center of mass energy of the
proton-proton collision at LHC is $14\,TeV$, black hole formation is
possible for $M_{min} < 14\,TeV$. So this figure shows the
possibility of formation and detection of $TeV$ black hole at the
LHC.}}
\end{figure}
\begin{figure}[htp]
\begin{center}
\includegraphics{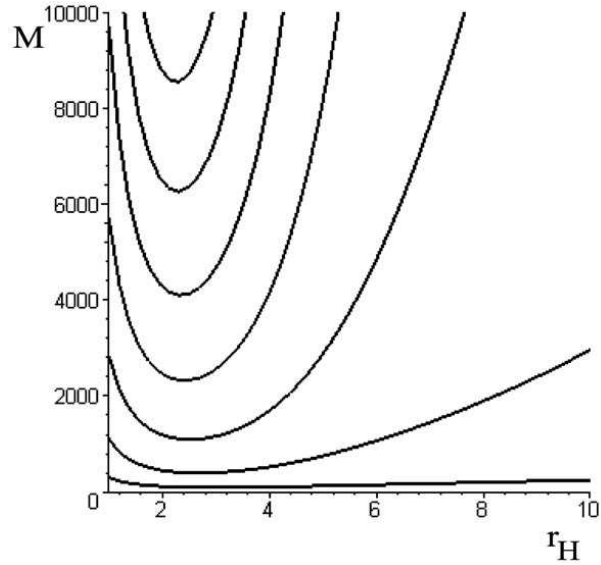}
\end{center}
\vspace{7 cm} \caption{\scriptsize {Black hole mass $M$ versus the
radius of event horizon $r_H$ for different number of spacetime
dimensions. Black hole charge and the $d$-dimensional Planck mass
are set equal to be $Q=0.5$ and $M_{Pl}=2$ respectively. On the
right-hand side of the figure, curves are marked from bottom to top
by $d = 4$ to $d = 10$. The figure shows that in this case there is
no black hole in the energy scales accessible at the LHC.}}
\end{figure}
As these two figures show, assuming a small enough $M_{Pl}$, it is
possible to detect the $TeV$ black holes at the expected
LHC-energies. The center of mass energy of the proton-proton ($pp$)
collision at LHC lab is $14\,TeV$. In this noncommutative framework
black hole formation is possible only for some minimum mass of
$M_{min} < 14\,TeV$. This is a pure noncommutative effect. In
commutative case this minimum value reduces to zero. As figure $4$
(with $M_{Pl}=0.4\,TeV$ and spacetime dimension  $d=6$) shows, the
minimum black hole mass in this situation is $1.8\, TeV$\,
approximately. In figure $5$, which is determined with
$M_{Pl}=2\,TeV$ and $d=6$, the minimum mass of the black hole is
going to be about $1100\, TeV$ which is impossible to be formed in
LHC. Hence, the possibility of forming these $10^3\,TeV$ black holes
at the LHC shrink to zero, however it is possible to be formed in
the ultrahigh energy cosmic ray (UHECR) airshowers [17]. Also,
figures $4$ and $5$ show that, if the number of spacetime dimension
increases at a small enough $M_{Pl}$, then the probability of
forming and producing black hole at the LHC will increase. On the
other hand, in this situation the minimal black hole mass threshold
for producing and detecting black hole at the LHC reduces. Contrary
to this, if the number of spacetime dimension, $d$, increases with a
larger amount of $d$-dimensional Planck mass, then the minimum
energy for black hole formation in collisions will increase and we
will not see any black hole at the usual $TeV$ energy scales.

The metric (12) shows a meaningful asymptotic behavior at short
distances. By using the asymptotic form of the metric (12), we find
the de Sitter type solutions with line element such that
\begin{equation}
 g_{00}= 1 -\frac{c_d\,M\,G_d}{\pi^{(\frac{d-3}{2})}\, \theta^{(\frac{d-1}{2})}}\, r^2
 +O\left(\,r^3\,\right),
\end{equation}
where $c_d$ is a dimensionless numerical constant which depends on
the number of spacetime dimensions. Since the physical effect of
noncommutativity is realized by substituting the position
Dirac-delta corresponding to point-like profiles with Gaussian
function of minimal width $\sqrt{\theta}$ describing the
corresponding smeared profiles [5,9,14,16,19], this form of
structure has a regular de Sitter vacuum solution accounting for the
effect of noncommutative coordinate fluctuations at short distances.
The effective density of vacuum energy corresponds to the effective
cosmological constant,
\begin{equation}\,\Lambda_{eff}=\frac{c_d\,M\,G_d}{\pi^{(\frac{d-3}{2})}\,
\theta^{(\frac{d-1}{2})}},
\end{equation}
which is leading to a finite curvature in the origin. It is
interesting to see that there is no {\it charge} term in the
effective cosmological constant. This is due to the fact that the
electric field has linear behavior at short distances [14], which
can only give raise to charge term of order $O\left(\, r^3\,\right)$
in the metric. Thus, an observer close to the origin sees only a
vacant mass $M$ without any charge contribution.\\
It is believed that noncommutativity can cure divergences that
appear, under the variety of forms, in General Relativity. For
instance, it would be of special interest to investigate the final
stage of black hole evaporation and some related thermodynamical
quantities of black hole in the framework of noncommutative
coordinates. In the next section we study this issue with details.

\section{Thermodynamics of Noncommutative RN-dS Black Holes}
Black hole thermodynamics has continued to fascinate researchers
since Hawking's discovery of the thermal radiation from black holes,
because it prepares a real connection between gravity and quantum
mechanics. The study of black hole thermodynamics also played a
crucial role in the extension of quantum field theory in curved
spacetime [20,21]. Hawking radiation shows how quantum fields on
black hole backgrounds behave thermally. In this regard, black hole
evaporation due to Hawking radiation is one of the fascinating
dynamical behaviors of a black hole structure. Although black holes
are perhaps the most perfectly thermal objects in the universe, but
their thermal properties are not fully understood yet. This section
aims to analyze some thermodynamical properties of the RN-dS black
hole and some problems about the final stage of black hole
evaporation in $d$-dimension with the hope that a little progress in
this direction to be achieved. Therefore, our next step is to
determine the thermodynamic behavior of noncommutative inspired
RN-dS black holes. To do this end, we should calculate Hawking
temperature of the black hole. The Hawking temperature can be
obtained in the usual manner by remembering that
\begin{equation}
T_H={1\over {4\pi}} {{dg_{00}}\over {dr}}|_{r=r_+}.
\end{equation}
When $d$ is odd, we can solve this equation analytically, however
for even $d$, it is impossible to solve it analytically and we must
perform numerical calculation of Hawking temperature. Black hole
temperature with some odd number of dimensions can be calculated as
follows
\begin{equation}
d=5\rightarrow
T_H=\frac{1}{4\pi}\Bigg(-MG_5\bigg[\frac{r_+}{3\pi\theta^2}+\frac{4X_5}
{3\pi\theta r_+^3}\bigg]+Q^2G_5\bigg[-\frac{\pi
X_5e^{-\frac{r_+^2}{4\theta}}}{12\theta^3r_+}-\frac{\pi
X_5^2}{3\theta^2r_+^5}\bigg]\Bigg),
\end{equation}

\begin{equation}
d=7\rightarrow
T_H=\frac{1}{4\pi}\Bigg(-MG_7\bigg[\frac{r_+e^{-\frac{r_+^2}{4\theta}}}{20\,\pi^2\theta^3}+\frac{2X_7}
{5\pi^2\theta^2r_+^5}\bigg]+Q^2G_7\bigg[-\frac{\pi
X_7e^{-\frac{r_+^2}{4\theta}}}{2560\,\theta^5r_+^3}-\frac{\pi
X_7^2}{320\,\theta^4r_+^9}\bigg]\Bigg),
\end{equation}

\begin{equation}
d=9\rightarrow
T_H=\frac{1}{4\pi}\Bigg(-MG_9\bigg[\frac{r_+e^{-\frac{r_+^2}{4\theta}}}{112\,\pi^3\theta^4}+\frac{3X_9}
{28\pi^3\theta^3r_+^7}\bigg]+Q^2G_9\bigg[-\frac{\pi
X_9e^{-\frac{r_+^2}{4\theta}}}{774144\,\theta^7r_+^5}-\frac{\pi
X_9^2}{64512\,\theta^6r_+^{13}}\bigg]\Bigg),
\end{equation}
and so on. \, $X_5$, $X_7$, and $X_9$ are functions of $r_+$ and
$\theta$ defined as follows
\begin{equation}
X_5=e^{-\frac{r_+^2}{4\theta}}\Big(r_+^2+4\theta-4\theta
e^{\frac{r_+^2}{4\theta}}\Big),
\end{equation}
\begin{equation}
X_7=e^{-\frac{r_+^2}{4\theta}}\Big(r_+^4+8\theta r_+^2+32\theta^2
-32\theta^2e^{\frac{r_+^2}{4\theta}}\Big),
\end{equation}
\begin{equation}
X_9=e^{-\frac{r_+^2}{4\theta}}\Big(r_+^6+12\theta r_+^4+96\theta^2
r_+^2+384\theta^3-384\theta^3 e^{\frac{r_+^2}{4\theta}}\Big).
\end{equation}
For even number of dimensions there are no closed analytical forms.
So, with numerical calculation of Hawking temperature in  arbitrary
number of spacetime dimensions, we show the results in forthcoming
figures. For simplicity, hereafter we set $\theta=1$ in numerical
calculations.

One motivation toward production and detection of micro-black holes
in collider tests is that their evaporation process is not so clear
for us. The evaporation process for charged black hole in the
framework of noncommutativity or the generalized uncertainty
principle [4,22] is usually arranged in two phases. In the former
phase, the temperature of the black hole grows during its
evaporation until it approaches to a maximum value which is
well-known to the Hawking phase. The latter phase is noncommutative
or GUP phase where in the noncommutative framework the temperature
suddenly falls down from Hawking phase maximum to zero [14] while in
the GUP framework it reaches to a nonzero, UV cutoff case with a
finite nonzero mass which is known as Planck size {\it remnant} [4].
Therefore, the evaporation process terminates when black hole mass
reaches to a fundamental mass and a consequent missing energy of
order $TeV$. The basic idea of a remnant is introduced by arguments
that to cure the information loss problem [3]. The formation of
stable black hole remnants would provide fascinating new signatures
which admit for the recognition of such a black hole remnant event
at near-future collider or UHECR experiments. Since the black hole
remnant carries a significant fraction of the total micro-black hole
mass, an amount of energy will be spent by the black hole in the
Hawking thermal radiation. When the evaporation process ends, the
black hole mass is in the Planck scale, leaving a remnant and an
effective missing energy can be observed by searching for events of
order $TeV$ missing energy. Also, charged black hole remnants could
remain a firm ionizing path electrically in the LHC detectors,
\textit{e.g.} ALICE, ATLAS, and
CMS, that this could let to recognize the black hole remnants.\\

\begin{figure}[htp]
\begin{center}
\includegraphics{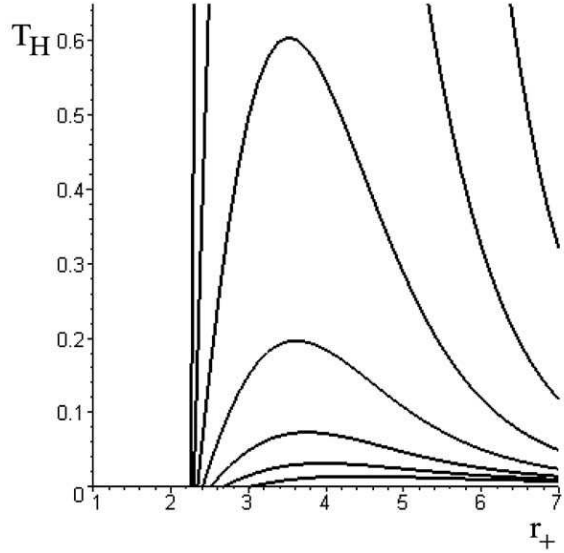}
\end{center}
\vspace{7 cm} \caption{\scriptsize {Black hole temperature, $T_H$,
as a function of $r_+$ for different number of spacetime dimensions.
In this figure, black hole charge, mass and the $d$-dimensional
Planck mass are set to be $Q=0.5$, $M=5$ and $M_{Pl}=0.4$,
respectively. On the right-hand side of the figure, curves are
marked from bottom to top by $d = 4$ to $d = 10$. Figure shows that
extra-dimensional black holes are hotter than four-dimensional black
holes on the recognized regime. }}
\end{figure}

\begin{figure}[htp]
\begin{center}
\includegraphics{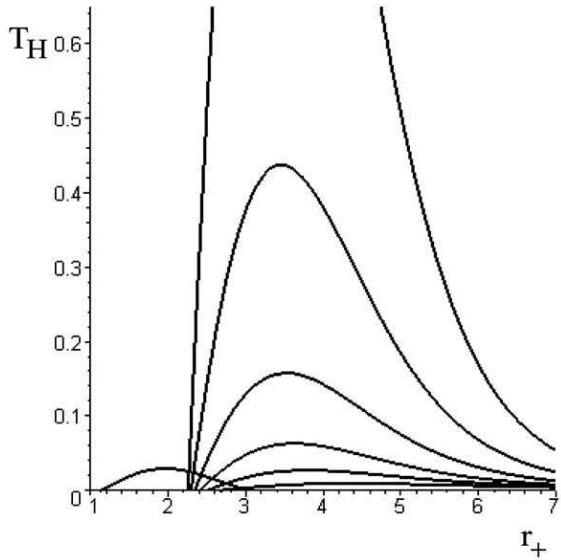}
\end{center}
\vspace{7 cm} \caption{\scriptsize {Black hole temperature, $T_H$,
as a function of $r_+$ for different number of spacetime dimensions.
Black hole charge and mass and the $d$-dimensional Planck mass are
set equal to $Q=2$, $M=5$ and $M_{Pl}=0.4$ respectively. On the
right-hand side of the figure, curves are marked from bottom to top
by $d = 4$ to $d = 10$. The figure shows that, when the black hole
charge varies  main changes will be occurred on the brane (the short
curve on the left-hand side of the figure). }}
\end{figure}

\begin{figure}[htp]
\begin{center}
\includegraphics{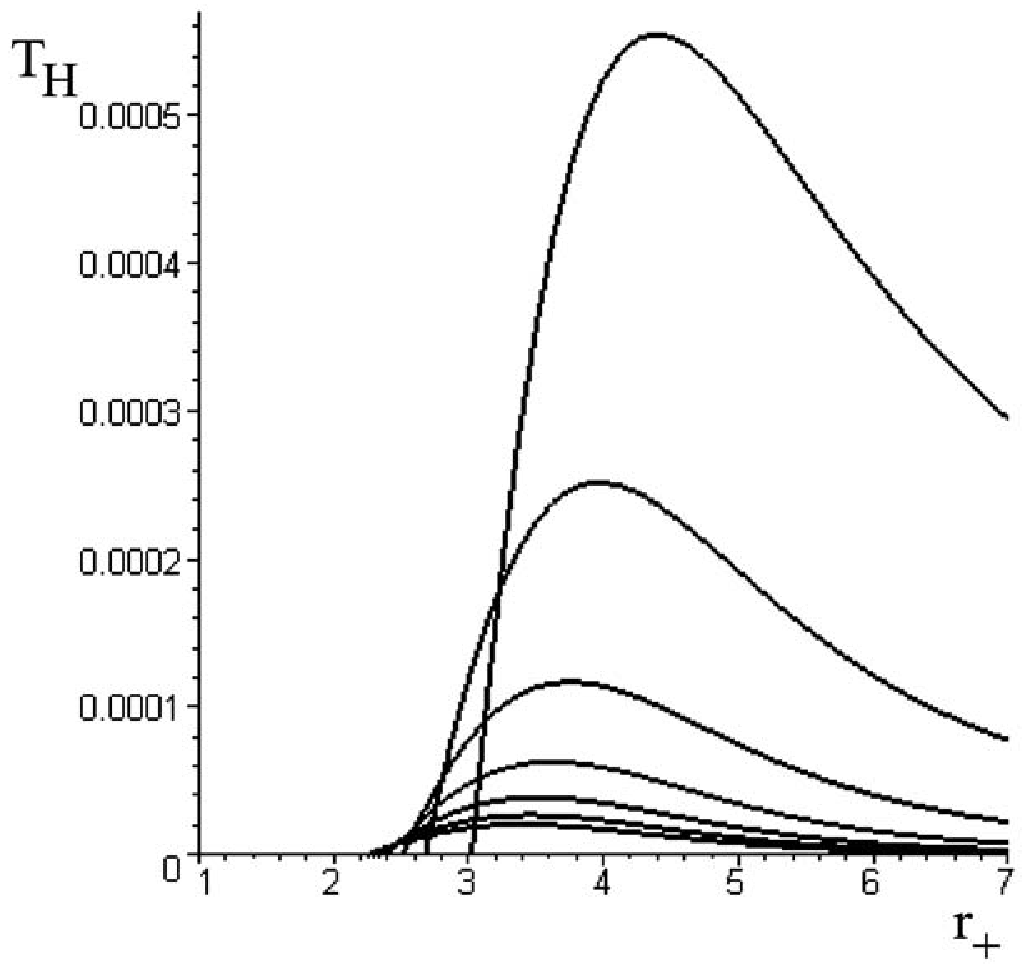}
\end{center}
\vspace{5.2 cm} \caption{\scriptsize {Black hole temperature, $T_H$,
as a function of $r_+$ for different number of spacetime dimensions.
Black hole charge and mass and the $d$-dimensional Planck mass are
set equal to $Q=0.5$, $M=5$ and $M_{Pl}=2$ respectively. On the
right-hand side of the figure, curves are marked from top to bottom
by $d = 4$ to $d = 10$. The figure shows that with this value of
$M_{Pl}$, contrary to figure $6$, the extra-dimensional black holes
are colder than four-dimensional black holes on the recognized
regime.}}
\end{figure}

As figure $6$ shows, assuming the fundamental Plank mass to be $
0.4\, TeV$, the Hawking temperature increases with increasing the
number of spacetime dimensions. Moreover the black hole remnant in
extra dimensions has smaller mass than 4-dimensional one. Therefore,
assuming a small enough fundamental energy-scales we expect
micro-black holes in higher-dimensional spacetime to be hotter, and
with a smaller mass at the endpoint of evaporation than
4-dimensional spacetime. When the charge of black hole varies as is
shown in figure $7$, increasing the charge leads to decreasing the
black hole temperature in a bulk spacetime but main changes occurs
on the 3-brane due to the fact that in LED scenarios, all
standard-model particles are limited to our observable 3-brane,
whereas gravitons can propagate the whole $d$-dimensional bulk
substantially. As Emparan {\it et al} have shown, the main energy
during Hawking radiation process from a $d$-dimensional black hole
is emitted within modes on the brane because there are a great
number of brane modes for standard model particles. Therefore, the
main energy is expected to be radiated on the brane but there is
only one graviton mode in the extra dimensions which can be
propagated in the bulk [18]. Moreover, the numerical result for
$d=4$ shows that no black hole is formed on the brane in this
region. Eventually, in figure $8$, by choosing $M_{Pl}=2\, TeV$, we
see that Hawking temperature decreases with increasing the number of
spacetime dimensions, however black hole remnants masses will be
smaller than $4$-dimensional counterpart as shown in previous
figures. Therefore, we expect micro-black holes in
higher-dimensional spacetime with a large fundamental energy-scale
to be colder, and again with a smaller mass remnant than
$4$-dimensional counterpart. Our inspection has shown that for
$M_{Pl}=1.155 \,TeV$, maximum Hawking temperature of black hole for
$d=10$ is approximately equal to Hawking temperature of $d=4$ black
hole. For $M_{Pl}> 1.155 \,TeV$ and $d\leq 10$, black holes in extra
dimensions are colder. Table $1$ shows these results. As a general
result, if large extra dimensions do really exist and the
$d$-dimensional Planck mass to be less than $1 TeV$, a great number
of black holes can be produced and detected in near-future
colliders.\\
\begin{table}
\caption{Comparison between black hole maximum temperature in four
and extra spacetime dimensions for different values of $M_{Pl}$.}
\begin{center}
\begin{tabular}{|c|c|}
\hline& $Q=0.5$ \, \, and \, \, $M=5\, TeV$   \\
\hline$M_{Pl}=0.911\, TeV$ & $T_H(max)|_{d=4}\approx T_H(max)|_{d=5}$ \\
\hline$M_{Pl}=0.915\, TeV$ & $T_H(max)|_{d=4}\approx T_H(max)|_{d=6}$ \\
\hline$M_{Pl}=0.966\, TeV$ & $T_H(max)|_{d=4}\approx T_H(max)|_{d=7}$ \\
\hline$M_{Pl}=1.026\, TeV$ & $T_H(max)|_{d=4}\approx T_H(max)|_{d=8}$\\
\hline$M_{Pl}=1.091\, TeV$ & $T_H(max)|_{d=4}\approx T_H(max)|_{d=9}$\\
\hline$M_{Pl}=1.155\, TeV$ & $T_H(max)|_{d=4}\approx T_H(max)|_{d=10}$\\
\hline
\end{tabular}
\end{center}
\end{table}

As another important thermodynamical properties, our next step is to
calculate and investigate status of entropy variations in such a
$d$-dimensional RN-dS black hole. This entropy is defined as
\begin{equation}
S=\int_{r_0}^{r_+}dr\,T_H^{-1}\,\frac{\partial M}{\partial r},
\end{equation}
where we find $S=0$ for the minimum horizon radius, $r=r_0$ (where
black hole mass is minimized), which is a reasonable choice. A
numerical evaluation of this expression for $M_{Pl}=2$ is shown in
figure $9$. The existence of the remnants is again approved from the
thermodynamical behavior of such a system.

\begin{figure}[htp]
\begin{center}
\includegraphics{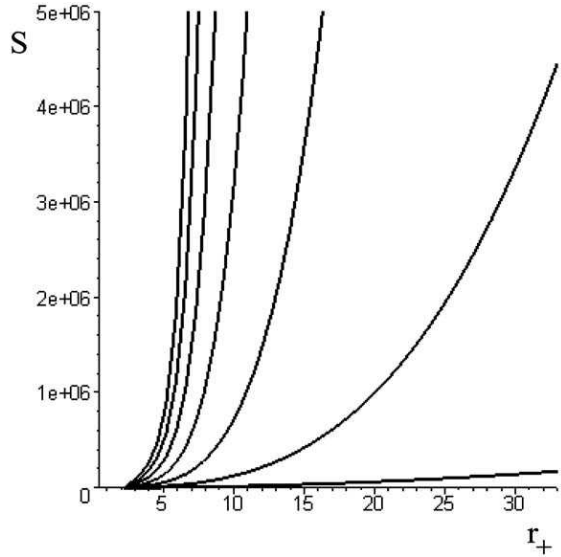}
\end{center}
\vspace{6.5 cm} \caption{\scriptsize {Black hole entropy, $S$, as a
function of $r_+$ for different number of spacetime dimensions.
Black hole charge and the $d$-dimensional Planck mass are set equal
to $Q=0.5$ and $M_{Pl}=2$ respectively. On the right-hand side of
the figure, curves are marked from bottom to top by $d = 4$ to $d =
10$. This figure shows that entropy increases with increasing the
number of extra dimensions. Note that this result is depended on the
value of $M_{Pl}$. For smaller values of $M_{Pl}$, the result is
completely different.}}
\end{figure}
Because of unusual thermodynamical properties of\,  $TeV$ black
holes in noncommutative scenarios, it is interesting to investigate
further thermodynamical details of these quantum gravity system. We
first study the heat capacity of the black hole which can be
obtained using the following relation
\begin{equation}
C=\frac{\partial M}{\partial r_+}\,\Bigg(\frac{\partial
T_H}{\partial r_+}\Bigg)^{-1}.
\end{equation}
The numerical results for $M_{Pl}=2$ is presented in figure $10$.
This figure shows that black hole has a negative heat capacity (it
means that $\Big(\frac{\partial T_H}{\partial r_+}\Big)^{-1}< 0$ \,
therefore \,  $ C<0$ in the Hawking phase) with a singularity for
fixed values of $Q$, $M_{Pl}$ and $d$. In fact, when the temperature
reaches a maximum value of its amount where the slope of temperature
curve $\Big(\frac{\partial T_H}{\partial r_+}\Big)^{-1}=0$ for a
special $r_+$ value, then the heat capacity becomes singular for
this special value of $r_+$. For lower $r_+$, the temperature falls
down (it means that $\Big(\frac{\partial T_H}{\partial
r_+}\Big)^{-1}> 0$\,\, gives \,\,$C>0$ \, in noncommutative or GUP
phase) to zero with a finite nonzero horizon radius, $r_0$ (which
means that $C=0$ for the final stage of black hole evaporation).

\begin{figure}[htp]
\begin{center}
\includegraphics{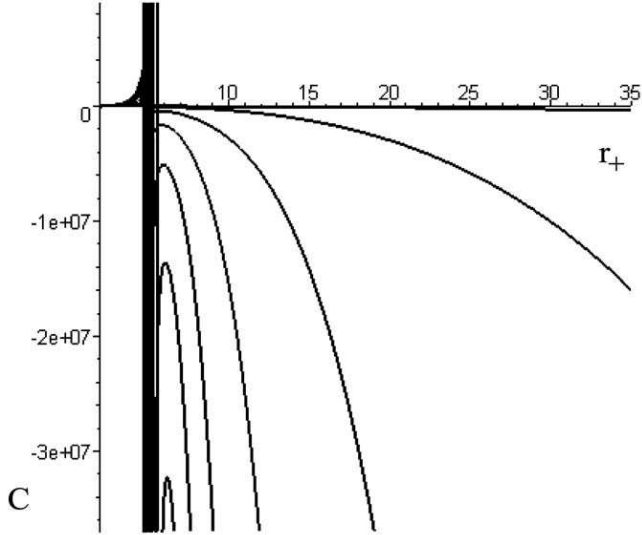}
\end{center}
\vspace{6.5 cm} \caption{\scriptsize {Black hole heat capacity, $C$,
as a function of $r_+$, for different number of spacetime
dimensions. Charge and the $d$-dimensional Planck mass are set equal
to $Q=0.5$ and $M_{Pl}=2$ respectively. On the right-hand side of
the figure, curves are marked from top to bottom by $d = 4$ to $d =
10$.}}
\end{figure}
Eventually, as a final remark in this section we consider the free
energy of the RN-dS black hole that can be defined as
\begin{equation}
F=M-T_H S.
\end{equation}
The numerical calculations of this quantity are presented in figures
$11$ and $12$. Reduction of the horizon size with decreasing free
energy and approaching negative values of free energy for large
values of $d$, can be seen both in these figures and in the equation
(28). It is evident that for $r=r_0$, the free energy becomes equal
to the minimum mass, $M_0$, due to the fact that temperature or
entropy are zero at this smeared-size, and therefore remnant is left
over.

\begin{figure}[htp]
\begin{center}
\includegraphics{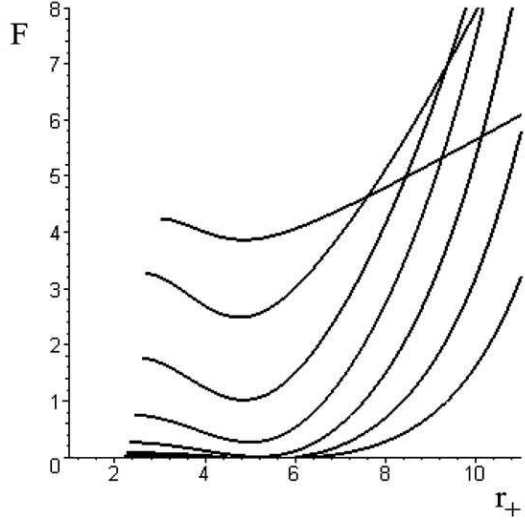}
\end{center}
\vspace{7 cm} \caption{\scriptsize {Black hole free energy, $F$, as
a function of $r_+$, for different number of spacetime dimensions.
Black hole charge and the $d$-dimensional Planck mass are set equal
to $Q=0.5$  and $M_{Pl}=0.4$ respectively. On the left-hand side of
the figure, curves are marked from top to bottom by $d = 4$ to $d =
10$. As is evident, the behavior of free energy in our 3-brane is
very different to other dimensions due to maximum effects of charge
on the 3-brane. The cutoff in the left hand side of the figure shows
the existence of remnant.}}
\end{figure}

\begin{figure}[htp]
\begin{center}
\includegraphics{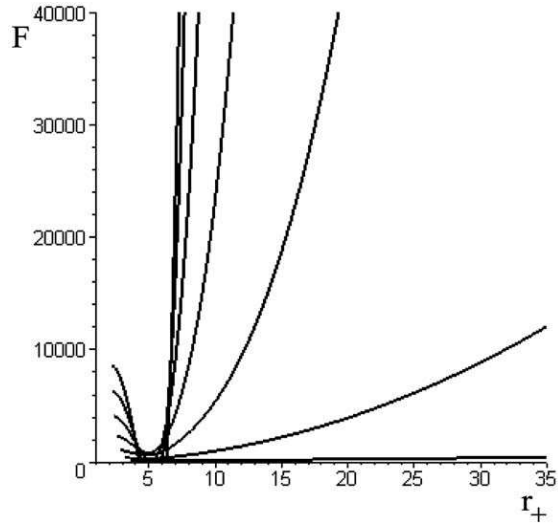}
\end{center}
\vspace{7 cm} \caption{\scriptsize {Black hole free energy, $F$, as
a function of $r_+$, for different number of spacetime dimensions.
Black hole charge and the $d$-dimensional Planck mass are chosen to
be $Q=0.5$ and $M_{Pl}=2$ respectively. On the right-hand side of
the figure, curves are marked from bottom to top by $d = 4$ to $d =
10$. Contrary to the previous figure, here we see that increasing
number of extra dimensions will increase the free energy of the
system for this value of $M_{Pl}$. The situation for small values of
$M_{Pl}$ is different as is shown in figure $11$.}}
\end{figure}

The idea of black hole remnant can cure both the singularity problem
at the endpoint of black hole evaporation and information loss
problem{\footnote{Recently, we have shown that the form of the
amendments for Hawking radiation as back-reaction effects with
incorporation of GUP influences can recover the information. In this
situation, correlations between the different modes of radiation
evolve, which reflect the fact that at least part of the information
leaks out from the black hole as the non-thermal GUP correlations
within the Hawking radiation [23].}. In fact, if a stable black hole
remnant is really exists due to the fact that there are some exact
continuous global symmetries in the nature [24], then the minimum
energy for black hole formation in collisions will be increased [25]
(but depending on the number of extra dimensions). In this
situation, the possibility of the production and detection of
$TeV$-scale black holes may be decreased because of lowering the
cross section for the expected LHC-energies and the absence of the
final decay particles for the detection in the LHC detectors, {\it
e.g.} ALICE, ATLAS, and CMS [26]. Therefore the idea of black hole
remnant is most meaningful for us. Of course, it is important to
note that if we consider the thermodynamic behavior at the very
short distances (mass scales smaller than minimal mass) then it
would be seen some exotic behavior of such a system. In a recent
paper [27] we have reported some results about extraordinary
thermodynamical behavior for Planck size black hole evaporation
which may reflect the need for a fractal nonextensive thermodynamics
[28] for Planck size black hole evaporation process. We just have
shown that if nothing halts the evaporation process, the
noncommutative black hole will throughout disappear eventually.
However, in this case one encounters some unusual thermodynamical
features leading to negative entropy, negative temperature and
anomalous heat capacity where the mass of the black hole becomes of
the order of Planck mass or less. There are two possible reasons for
these unusual features: either we really cannot trust the details of
the noncommutative effects with the Gaussian, Lorentzian and some
other profiles of the smeared mass distribution at the regions that
the mass of the black hole to be of the order of Planck mass [9]
(see also [29] and [30]), or we really should doubt the results of
standard thermodynamics at quantum gravity level which the origin of
this premise may possibly becomes due to the fractal nature of
spacetime at very short distances [27]. Indeed, at present we don't
know which of these ideas are true.

\section{Summary and Discussion}
The noncommutative version of quantum field theories based on Moyal
$\star$-product [31] lead to failure in resolving of some important
problems, such as Lorentz invariance breaking, loss of unitarity and
UV divergences of quantum field theory. Unfortunately, no flawless
and completely convincing theory of noncommutativity yet exists.
However, the authors in Ref.~[16] explained that the coordinate
coherent states approach as a fascinating model of noncommutativity
can be free from the problems mentioned above. In this approach,
General Relativity in its usual commutative manner as described by
the Einstein-Hilbert action remains applicable inasmuch, if
noncommutativity effects can be treated in a perturbative manner,
then this is defensible, at least to a good approximation. Indeed,
the authors in Ref.~[32] have shown that the leading
noncommutativity corrections to the form of the Einstein-Hilbert
action are at least second order in the noncommutativity parameter
$\theta$. The generalization of the quantum field theory by
noncommutativity based on coordinate coherent state formalism is
also interestingly curing the short distance behavior of pointlike
structures. Therefore, noncommutativity brings prominent qualitative
and quantitative changes to the properties of black hole
thermodynamics. Indeed, these changes could have important concepts
for the possible formation and detection of black holes at the
expected LHC-energies. In this paper, we have generalized the ANSS
model of noncommutative Reissner-Nordstr\"{o}m like geometries to
model universes with large extra dimensions. Noncommutativity
eliminates spacetime singularity due to smeared picture of particle
mass and charge. The energy scales for production and detection of
black holes remnants at LHC are examined and it has been shown that
in the presence of noncommutativity, thermodynamical properties of
$TeV$ black holes depend on the values of fundamental Planck mass in
extra dimensions. The possibility of black hole formation is reduced
by increasing the charge of black hole particularly for
$4$-dimensional black hole on the brane. Since the center of mass
energy of the proton-proton collision at LHC is $14\,TeV$, black
hole formation is possible for $M_{min} < 14\,TeV$. Our analysis
shows that if the number of spacetime dimension, $d$, increases with
a larger amount of $d$-dimensional fundamental Planck mass, then the
minimum energy for black hole formation in collisions will increase
and we will not see any black hole at the usual $TeV$ energy scales
at LHC. In contrast, a smaller amount of $d$-dimensional fundamental
Planck mass leads to conclusion that the minimum energy for black
hole formation in collisions will decrease with increasing the
number of extra dimensions and we are able to see black hole at the
usual $TeV$ energy scales at the LHC. We have obtained an effective
and noncommutative inspired cosmological constant in $d$-dimension
which is leading to a finite curvature in the origin. From
thermodynamics point of view, for a suitable choice of fundamental
mass scale, Hawking temperature increases with increasing the number
of spacetime dimensions. Moreover, the black hole remnant in extra
dimensions has smaller mass than 4-dimensional one. Assuming a small
enough fundamental energy-scales we expect micro-black holes in
higher-dimensional spacetime to be hotter, and with a smaller mass
at the endpoint of evaporation than 4-dimensional spacetime. When
the charge of black hole varies, increasing the charge leads to
decreasing the black hole temperature in a bulk spacetime but main
changes occurs on the 3-brane due to the fact that in LEDs
scenarios, all standard-model particles are limited to our
observable 3-brane, whereas gravitons can propagate the whole
$d$-dimensional bulk substantially. The situation for the case with
higher fundamental mass scale is different; in this situation the
extra-dimensional black holes are colder than four-dimensional black
holes on the recognized regime. Our analysis on $TeV$ black hole
production at the LHC shows that if large extra dimensions do really
exist and the $d$-dimensional Planck mass to be less than $1\, TeV$,
a great number of black holes can be produced and detected in LHC
and other near-future colliders.

As a remark we accentuate that some authors have presented the black
hole thermodynamics in the noncommutative framework adapting a
coordinate noncommutativity against coherent state approach (see [8]
and references therein). A question then appears: what is the
difference between these two approaches? The standard way to handle
the noncommutative problems is through the utilize of Moyal
$\star$-product. That means to use complex number commuting
coordinates and shift noncommutativity in the product between
functions. This is mathematically valid, but it is physically
useless since any model written in terms of $\star$-product, even
the simplest field theory, is nonlocal and it is not obvious how to
handle nonlocal quantum field theory. One suggested approach is
perturbation in the $\theta$ parameter [33]. This is physically
reasonable due to the fact that once expanded up to a given order in
$\theta$, the resulting field theory becomes local. The smeared
picture of particles based on coordinate coherent states defines
complex number coordinates as quantum mean values of the original
noncommuting ones between coordinate coherent states. In other
words, in this setup one can see commuting coordinates as classical
limit (in the quantum mechanical sense) of the noncommuting ones. In
this framework, the emergent semiclassical geometry keeps memory of
its origin. For example, free propagation of a point-like object is
described by a minimal width Gaussian wave-packet as has been
considered in our setup. So, the difference between two approaches
lies in the definition of quantum field theoretical propagators.\\

{\bf Acknowledgment}\\
This work has been supported partially by Research Institute for
Astronomy and Astrophysics of Maragha, Iran.\vspace{1cm}\\

\end{document}